\documentclass[twocolumn,pre,floatfix,showpacs]{revtex4}

\usepackage{epsfig}
\usepackage{amsmath}
\usepackage{times}

\usepackage{ifthen}

\def\be{\begin{equation}}
\def\ee{\end{equation}}
\def\ep{\epsilon}
\def\te{\tilde{\epsilon}}

\begin{document}
\bibliographystyle{revtex}
 
\title{Dielectrophoresis of charged colloidal suspensions}
\date{\today}

\author{J. P. Huang}
\affiliation{Biophysics and Statistical Mechanics Group,
Laboratory of Computational Engineering, Helsinki University of Technology, 
 P.\,O. Box 9203, FIN-02015 HUT, Finland, and \\
Department of Physics, The Chinese University of Hong Kong,
 Shatin, NT, Hong Kong}

\author{Mikko Karttunen}
\affiliation{Biophysics and Statistical Mechanics Group,
Laboratory of Computational Engineering, Helsinki University of Technology, 
 P.\,O. Box 9203, FIN-02015 HUT, Finland}

\author{K. W. Yu, and L. Dong}
\affiliation{
Department of Physics, The Chinese University of Hong Kong,
 Shatin, NT, Hong Kong}

\begin{abstract}

We present a theoretical study of dielectrophoretic (DEP) crossover
spectrum of two polarizable particles under the action of a nonuniform AC
electric field. For two approaching particles, the mutual polarization
interaction yields a change in their respective dipole moments, 
and hence, in the DEP crossover spectrum.
The induced polarization effects are captured by the multiple image
method. Using spectral representation theory, an analytic
expression for the DEP force is derived. We find that the mutual
polarization effects can change the crossover frequency at which the DEP
force changes sign. 
The results are found to be in agreement with recent experimental observation 
and as they go beyond the standard theory, they help to clarify the important 
question of the underlying polarization mechanisms.

\end{abstract}

\pacs{ 82.70.-y, 77.22.GM, 61.20.Qg, 77.84.Nh}

\maketitle

\section{Introduction}

When a polarizable particle is subjected to 
an applied electric field,
a dipole moment is induced into it. The movement of colloidal
particles in an applied AC electric field is called
dielectrophoresis~\cite{Pohl}.  It is typically used for
micromanipulation and separation of biological cellular size
particles, and it has recently been successfully applied to submicron
size particles as well.  Specific applications include diverse
problems in medicine, colloidal science and nanotechnology, e.g.
separation of nanowires~\cite{Duan}, viruses~\cite{Hug1}, latex
spheres~\cite{Mor1,marquet02a}, DNA~\cite{Chou} and leukemic
cells~\cite{ratanachoo02a}, as well as lab-on-a-chip designs~\cite{yhuang02a}.

The dielectrophoretic (DEP) force exerted on a particle can be either
attractive or repulsive depending on the polarizability of the
particle in comparison to the medium.  For a nonuniform AC electric
field, the magnitude and the direction of the DEP force depends on the
frequency, changes in surface charge-density and free charges in the
vicinity of the particle.  The frequency at which the DEP force
changes its sign is called the crossover frequency
($f_{\mathrm{CF}}$). Analysis of the crossover frequency as a function
of the host medium conductivity can be used to characterize the
dielectric properties of particles, and is at present the principal
method of DEP analysis for submicrometer particles~\cite{Hug1,Mor1}.

In the dilute limit, i.e, when a small volume fraction of charged
particles are suspended in an aqueous electrolyte solution, one can
focus on the DEP spectrum of an individual particle  ignoring the
mutual interactions between the particles. Although the current
theory~\cite{Pohl} captures some of the essential physics in the
dilute case, it is not adequate~\cite{JPCB,Bay,JCIS,khusid96a,green99a}. 
This is due to the fact that even for a single colloidal particle in an
electrolyte, it is not established which mechanisms control its
dielectric properties.  If the suspension is not dilute, the situation
becomes even more complicated due to the mutual interactions between
the particles.  One should also note that particles may aggregate due
to the presence of an external field, even when the suspension is at
the dilute limit under zero field conditions. In this case, the mutual
interactions have to be included in the description.

In this article, we present a theoretical study of the DEP spectrum of
two spherical particles in the presence of a nonuniform AC electric
field. We use the multiple image method~\cite{Yu00}, which is able to
capture the mutual polarization effects. Using the spectral
representation theory~\cite{Berg}, we derive an analytic expression
for the DEP force and determine the crossover frequency. Our
theoretical analysis shows that the induced mutual polarization
interactions plays an important role in DEP spectrum.  In a more
general framework, our results demonstrate the importance of
correlation effects. This is analogous to the findings in charged
systems where phenomena such as overcharging, or charge inversion (see
e.g. Refs.~\cite{grosberg:02a,patra02a}), provide spectacular
demonstrations of correlation effects.

As our starting point, we consider a pair of interacting charged
colloidal particles dispersed in an electrolyte solution. When the two
particles approach each other, the mutual polarization interaction
between them leads to changes in their respective dipole
moments~\cite{PRE2}, and hence also in the DEP spectrum and crossover
frequency.  We analyze two cases: 1) longitudinal field (L), in which
the field is parallel to the line joining the centers of particles,
and 2) transverse field (T) in which the field is perpendicular. The
former corresponds to positive dielectrophoresis where a particle is
attracted to regions of high field and the latter to the opposite
case, referred to as negative dielectrophoresis.

This paper is organized as follows. In Sec.~\ref{sec:forma} we present
the formalism and derive analytic expressions for the effective dipole
factors in spectral representation. In Sec.~\ref{sec:numer}, we use
the analytical results to numerically solve the crossover frequency,
dispersion strength and DEP spectra under different conditions. This
is followed by a discussion of the results in  Sec.~\ref{sec:concl}.

\section{Formalism and analysis}
\label{sec:forma}

First, we consider a single charged spherical particle suspended in an
electrolyte and subjected to a nonuniform AC electric field. The DEP
force ${\bf F}_{\mathrm{DEP}}$ acting on the particle is then given
by~\cite{Jones}
\be
{\bf F}_{\mathrm{DEP}}=
\frac{1}{4}\pi \ep_2D^3\mathrm{Re}[b]\nabla |{\bf E}|^2,
\label{depf}
\ee
where $D$ is particle diameter, $\ep_2$ the real dielectric constant
of host medium, ${\bf E}$ the local RMS electric field, and
$\mathrm{Re}[b]$ the real part of the dipole factor (also called
Clausius-Mossotti factor)
\be
b=\frac{\te_1-\te_2}{\te_1+2\te_2}.
\ee
Here, $\te_1$ and $\te_2$ are the complex dielectric constants of the
particle and the host medium, respectively. In order for the two above
equations to be valid in an AC field, the dielectric constant must
include dependence on the frequency.  The complex frequency dependent
dielectric constant is defined as $$
\te=\ep+ \frac{\sigma}{i2\pi f},
$$ where $\ep$ is the  real dielectric constant, $\sigma$ denotes
conductivity, $f$ the frequency of the external field, and $i \equiv
\sqrt{-1}$.

The conductivity of a particle consists of three components: Its bulk
conductivity ($\sigma_{\mathrm{1bulk}}$), surface effects due to the
movement of charge in the diffuse double layer (conductance $k_d$),
and the Stern layer conductance ($k_s$), i.e.,
\be
\sigma_1=\sigma_{\mathrm{1bulk}}+\frac{4k_d}{D}+\frac{4k_s}{D}.
\ee
The diffuse double layer conductance $k_d$ can be given as~\cite{Ly}
\be
k_d=\frac{4F_a^2cz^2 \Xi (1+3\Lambda/z^2)}
{R_0T_0\kappa}\left[\cosh\left(\frac{zF_a\zeta}{2R_0T_0}\right)-1\right],
\label{kd}
\ee
where $\Xi$ is the ion diffusion coefficient, $z$ the valency of
counterions, $F_a$ the Faraday constant, $R_0$ the molar gas constant,
$\zeta$ the electrostatic potential at the boundary of the slip plane
and $T_0$ the temperature. The reciprocal Debye length $\kappa$
providing a measure for screening on the system is given by
\be
\kappa=\sqrt{\frac{2czF_a^2}{\ep_2R_0T_0}},
\ee
where $c$ is the electrolyte concentration. Parameter
$\Lambda$ in Eq.~(\ref{kd}) describes the electro-osmotic contribution
to $k_d$, and it is given by
\be
\Lambda=\left(\frac{R_0T_0}{F_a}\right)^2\frac{2\ep_2}{3\eta \Xi},
\ee
where $\eta$ is the viscosity of medium. In addition, the Stern layer
conductance $k_s$ has the form~\cite{JCIS}
\be
k_s=\frac{u\mu_r \Sigma}{2zF_a},
\ee
where $u$ is the surface charge density, $\Sigma$ molar conductivity
for a given electrolyte, and $\mu_r$ gives the ratio between the ion
mobility  in the Stern layer to that in the medium.

For a pair of  particles at a  separation $R$ suspended in an
electrolyte, we have to consider the multiple image effect. We
consider two spheres in a medium, and apply a uniform electric field
${\bf E}_0=E_0\hat{z}$ to the suspension. This induces a dipole moment
into each of the particles.  The dipole moments of particles 1 and 2
are given by $p_{10}$ and $p_{20}(\equiv p_{10}=\ep_2E_0D^3b/8)$,
respectively.

Next, we include the image effects. The dipole moment $p_{10}$ induces
an image dipole $p_{11}$ into sphere 2, while $p_{11}$ induces another
image dipole in sphere 1. As a result, multiple images are
formed. Similarly, $p_{20}$ induces an image $p_{21}$ into colloid 1.
The formation of multiple images leads to an infinite series of image
dipoles.

In the following, we obtain the sum of  dipole moments inside each
particle, and derive the desired expressions for dipole factors.  We
consider two basic cases:  1) longitudinal field (L), where the field
is parallel to the line joining the centers of particles, and 2)
transverse field (T), where the field is perpendicular to the line
joining the centers of particles. Using the above notation, the
effective dipole factors for a pair are given by~\cite{Yu00}
\begin{eqnarray}
b_L{}^*&= &b\sum_{n=0}^{\infty}(2b)^n\left[\frac{\sinh \alpha}{\sinh
(n+1)\alpha}\right]^3,\nonumber\\ b_T{}^*&=
&b\sum_{n=0}^{\infty}(-b)^n\left[\frac{\sinh \alpha}{\sinh
(n+1)\alpha}\right]^3,
\label{eq:blbt}
\end{eqnarray}
where $\alpha$ is defined via  $\cosh\alpha=R/D$. The summations  in
Eqs.~(\ref{eq:blbt}) include the multiple image effects,  the $n=0$
term giving the dipole factor of an isolated particle.

We have to derive the analytic expressions for $\mathrm{Re}[b_L^*]$
and $\mathrm{Re}[b_T^*]$ to resolve the DEP force in
Eq.~(\ref{depf}). To do that, we resort to spectral representation
theory. It offers the advantage of being able to separate the material
parameters (such as dielectric constant and conductivity)  from
structural information~\cite{Berg} in a natural way.

Let us begin by defining a complex material parameter
$\tilde{s}=1/(1-\te_1/\te_2)$.  Using this, the dipole factor for a
pair takes the form
\be
b^*=\sum_{n=1}^{\infty}\frac{F_n}{\tilde{s}-s_n},
\ee
where $n$ is a positive integer, and $F_n$ and $s_n$ are the $n-$th
microstructure parameters of the composite material~\cite{Berg}. As an
example, the dipole factor of an isolated particle in spectral
representation expression becomes $b=F_1/(\tilde{s}-s_1)$, where
$F_1=-1/3$ and $s_1=1/3$.

In order to obtain expressions for the dipole factors $b_L^*$ and
$b_T^*$ in Eqs.~(\ref{eq:blbt}), we introduce the following identity $$
\frac{1}{\sinh^3x}=\sum_{m=1}^{\infty}4m(m+1)\exp[-(1+2m)x].
$$ Its application into Eqs.~(\ref{eq:blbt}) yields the following
exact transformations:
\begin{eqnarray}
b_L^*&=&\sum_{m=1}^{\infty}\frac{F_m^{(L)}}{\tilde{s}-s_m^{(L)}},\nonumber\\
b_T^*&=&\sum_{m=1}^{\infty}\frac{F_m^{(T)}}{\tilde{s}-s_m^{(T)}},
\label{eq:bspec}
\end{eqnarray}
where the m-th components of the microstructure parameter of 
the composite material are given as
\begin{eqnarray}
F_m^{(L)}&\equiv & F_m^{(T)}=-{4 \over 3}m(m+1)\sinh^3\alpha
\exp[-(2m+1)\alpha],\nonumber\\ s_m^{(L)}&=&{1\over
3}\{1-2\exp[-(1+2m)\alpha]\},\nonumber\\ s_m^{(T)}&=&{1\over
3}\{1+\exp[-(1+2m)\alpha]\}.\nonumber
\end{eqnarray}

To make this approach more tractable, we introduce dimensionless
dielectric constant and conductivity~\cite{Lei}, $s=1/(1-\ep_1/\ep_2)$
and $t=1/(1-\sigma_1/\sigma_2)$, respectively. Now, we are able
separate the real and imaginary parts of the arguments in the
expressions for  $b_L^*$ and $b_T^*$ in Eq.~(\ref{eq:bspec}).  The
argument can be rewritten as
\be
\frac{F_m}{\tilde{s}-s_m}=(\frac{F_m}{s-s_m}
+\frac{\Delta\ep_m}{1+f^2/f_{mc}^2}) -i\frac{\Delta\ep_m
f/f_{mc}}{1+f^2/f_{mc}^2}
\label{eq:fm}
\ee
where
\be
\Delta\ep_m=F_m\frac{s-t}{(t-s_m)(s-s_m)}
\label{eq:epm}
\ee
and
\be
\ f_{mc}=\frac{1}{2\pi}\frac{\sigma_2 s(t-s_m)}{\ep_2 t(s-s_m)}.
\label{eq:fmc}
\ee
The analytic expressions for $\mathrm{Re}[b_L^*]$ and $\mathrm{Re}[b_T^*]$
(Eq.~(\ref{eq:bspec})) become
\begin{eqnarray}
\mathrm{Re}[b_L^*]&=&\sum_{m=1}^{\infty}(\frac{F_m^{(L)}}{s-s_m^{(L)}}
+\frac{\Delta\ep_m^{(L)}}{1+f^2/f_{mc}^{2(L)}}),\nonumber\\
\mathrm{Re}[b_T^*]&=&\sum_{m=1}^{\infty}(\frac{F_m^{(T)}}{s-s_m^{(T)}}
+\frac{\Delta\ep_m^{(T)}}{1+f^2/f_{mc}^{2(T)}}).
\end{eqnarray}
Using these, we can obtain the DEP force ${\bf F}_{\mathrm{DEP}}$
which  includes corrections due to the image effects. The DEP spectrum
consists of a series of sub-dispersions with strength $\Delta\ep_m$
and characteristic frequency $f_{\mathrm{mc}}$. In particular, the
frequency which yields $F=0$, namely $\mathrm{Re}[b^*]=0$, is the desired
crossover frequency $f_{\mathrm{CF}}$.

\section{Numerical results}   
\label{sec:numer}

The above formalism enables us to study the effects 
due to multiple images under different physical conditions and to
compare the theory to experimental results. In the following, 
we compare the crossover frequency of an isolated particle to that of
two particles at different separations. We study the effects due to 
multiple images
by varying medium conductivity, the $\zeta$-potential, medium viscosity,
surface charge density, real dielectric constant of the particle and
molar conductivity. Finally, we have computed the DEP spectrum and 
the dispersion strength. 

\begin{figure}
\centering
\hspace*{-1cm}\epsfig{file=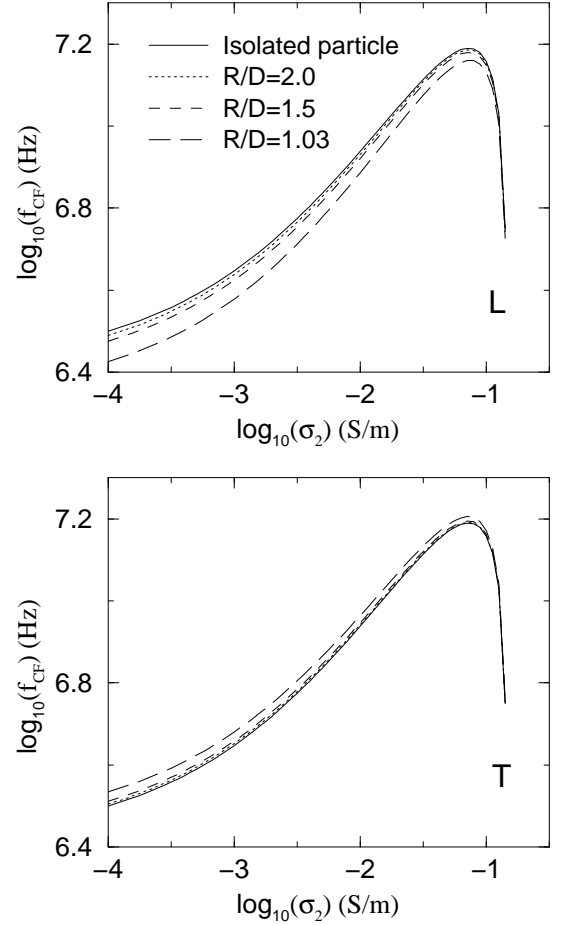,width=6.2in}
\caption{DEP crossover frequency vs. medium conductivity for 
an isolated particle (solid line) and two particles at different separations. 
L denotes longitudinal field case and T transverse field case. 
Parameters: $\zeta=0.12\,$ V, $\eta=1.0\times 10^{-3}\,$Kg/(ms), 
$u=0.033\,$C/m$^2$, $\Sigma=0.014\,$Sm$^2$/mol, $\ep_1=2.25\ep_0$. }
\label{Fig.1.}
\end{figure}

The common parameters used in all numerical computations are the following: 
Temperature $T_0=293\,$K, dielectric constant
of host medium $\ep_2=78\ep_0$, 
bulk conductivity of the colloidal particle
$\sigma_{1bulk}=2.8\times 10^{-4}\,$S/m, 
ion diffusion coefficient $\Xi=2.5\times 10^{-9}\,$m$^2$/s, 
the ratio between the ion
mobility  in the Stern layer to that in the medium $\mu_r=0.35$, 
particle diameter $D=2.16\times 10^{-7}$m, counterion valency $z=1$. 
The dielectric constant of vacuum is denoted by $\ep_0$.

Figure~\ref{Fig.1.} shows the DEP crossover frequency as a function of 
medium conductivity for an isolated particle and for two particles at
different separations.
In agreement with recent experiments~\cite{Hug2}, 
we find that a peak in 
the crossover frequency appears at a certain medium conductivity.
The appearance of a peak is preceded by an increase
of  $f_{\mathrm{CF}}$ upon increasing medium conductivity and 
followed by an abrupt drop~\cite{JCIS, Hug2}.
Compared to an isolated particle, the multiple image effect leads to a  
red-shift (blue-shift)  in $f_{\mathrm{CF}}$ in the longitudinal (transverse)
field case. Furthermore, for longitudinal (transverse) field, 
the stronger the polarization interaction, the lower (higher) the crossover frequency.
In addition, it is worth noting that the effect of the multiple 
images is the opposite in the longitudinal  and transverse cases.
As  the ratio $R/D$ grows, the predicted crossover spectrum approaches 
to that of an isolated particle, i.e., at large separations 
the multiple image interaction becomes negligible. 
\begin{figure}
\centering
\hspace*{-1cm}\epsfig{file=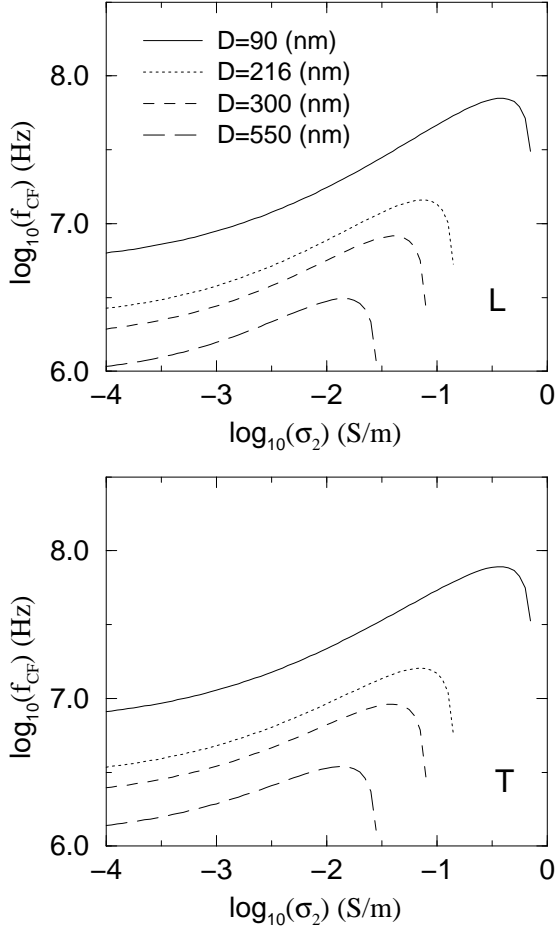,width=6.2in}
\caption{DEP crossover frequency vs. medium conductivity when
the particle size is varied. 
Parameters as in Fig.~\ref{Fig.1.}.
}
\label{Fig.1b.}
\end{figure}

Motivated by a recent experiment~\cite{green99a}, 
we analyzed the effect of particle size on 
the crossover frequency by 
keeping the ratio $R/D$ fixed and varying the particle diameter.
In agreement with the experiments, 
we find that the location of the peak is shifted to higher frequencies 
and higher conductivities when the diameter
of the particle is reduced, see Fig.~\ref{Fig.1b.}. 

\begin{figure}
\centering
\hspace*{-1cm}\epsfig{file=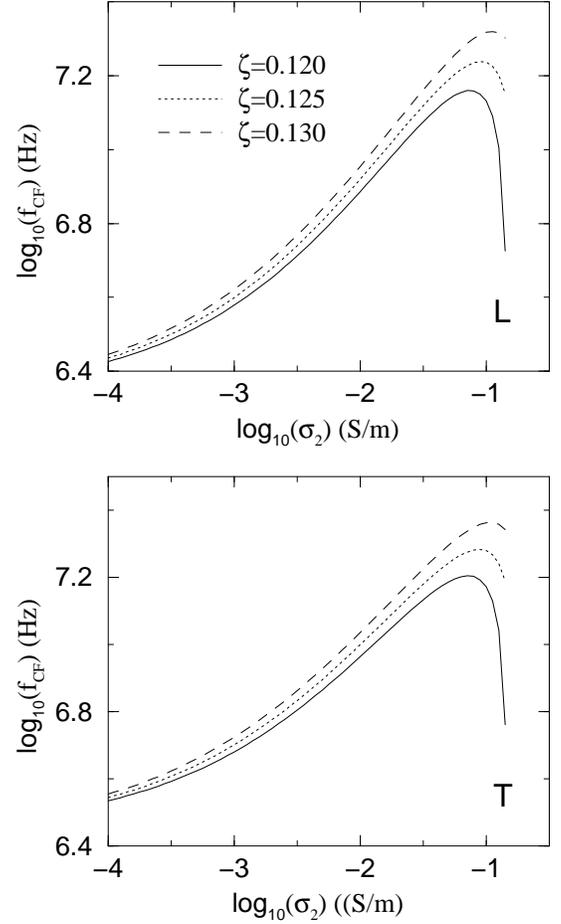,width=6.2in}
\caption{DEP crossover frequency vs. medium conductivity
for different $\zeta$-potentials. 
Parameters as in Fig.~\ref{Fig.1.}.
}
\label{Fig.2.}
\end{figure}

\begin{figure}
\centering
\hspace*{-1cm}\epsfig{file=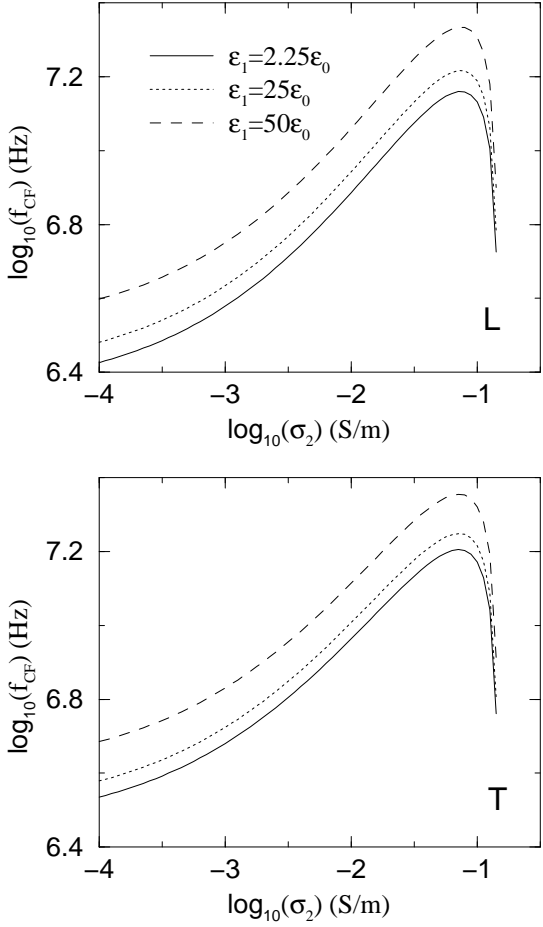,width=6.2in}
\caption{DEP crossover frequency vs. medium conductivity.
The real part of the dielectric constant is varied. 
Parameters as in Fig.~\ref{Fig.1.}.
}
\label{Fig.5.}
\end{figure}

\begin{figure}
\centering
\hspace*{-1cm}\epsfig{file=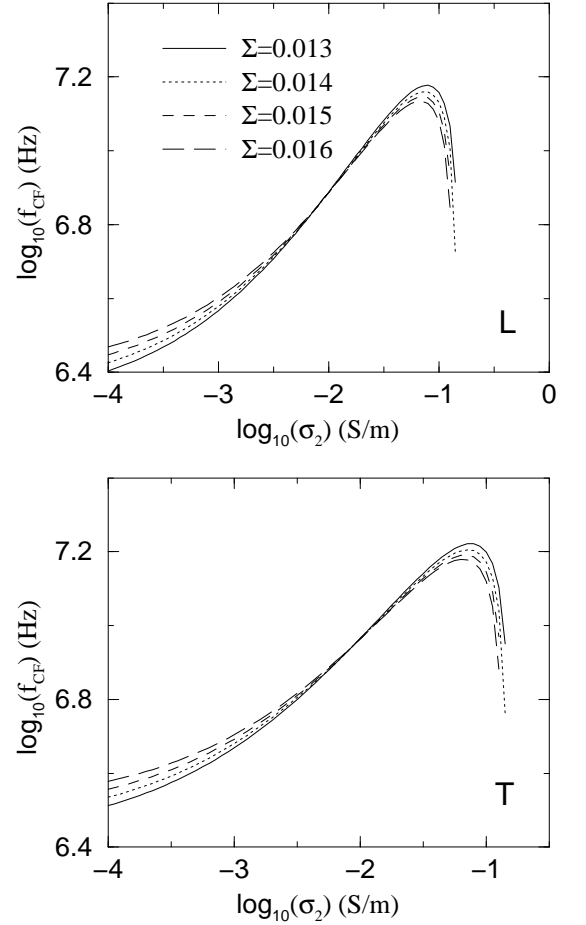,width=6.2in}
\caption{The effect of molar conductivity on the DEP crossover frequency. 
Parameters as in Fig.~\ref{Fig.1.}.
}
\label{Fig.6.}
\end{figure}

\begin{figure}
\centering
\hspace*{-1cm}\epsfig{file=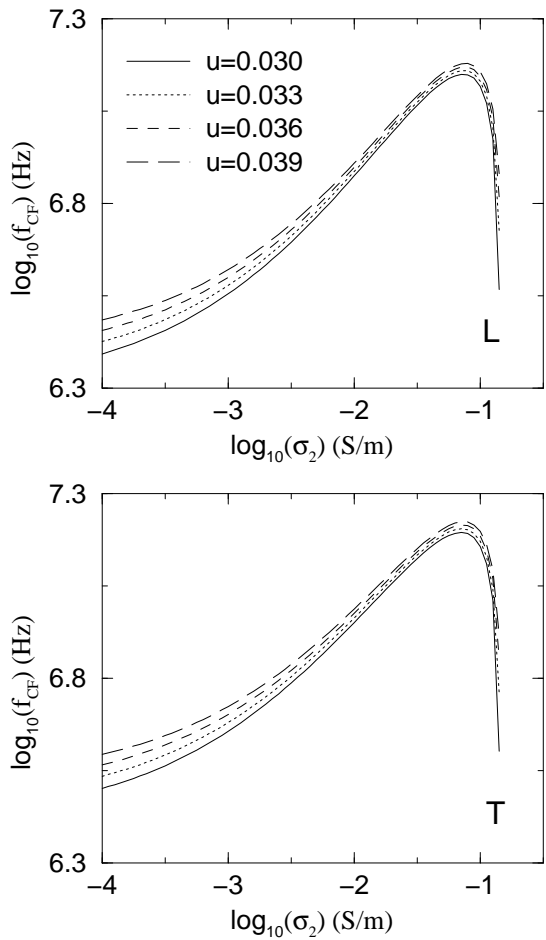,width=6.2in}
\caption{The effect of surface charge density on the DEP crossover frequency.
Parameters as in Fig.~\ref{Fig.1.}.
}
\label{Fig.4.}
\end{figure}

\begin{figure}
\centering
\hspace*{-1cm}\epsfig{file=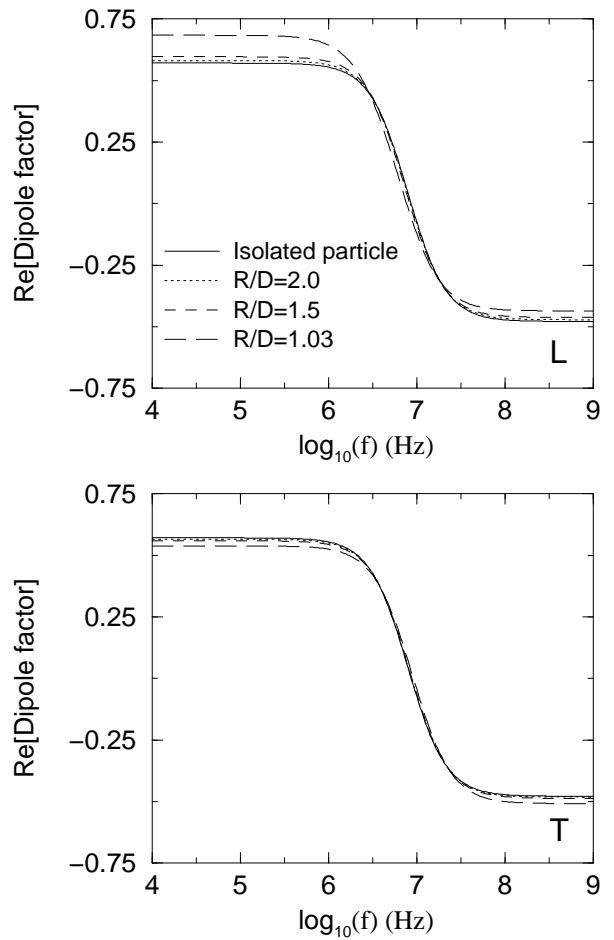,width=6.2in}
\caption{DEP spectrum (the real part of the dipole factor). 
Parameters as in Fig.~\ref{Fig.1.}.
}
\label{Fig.7.}
\end{figure}

\begin{figure}
\centering
\hspace*{-1cm}\epsfig{file=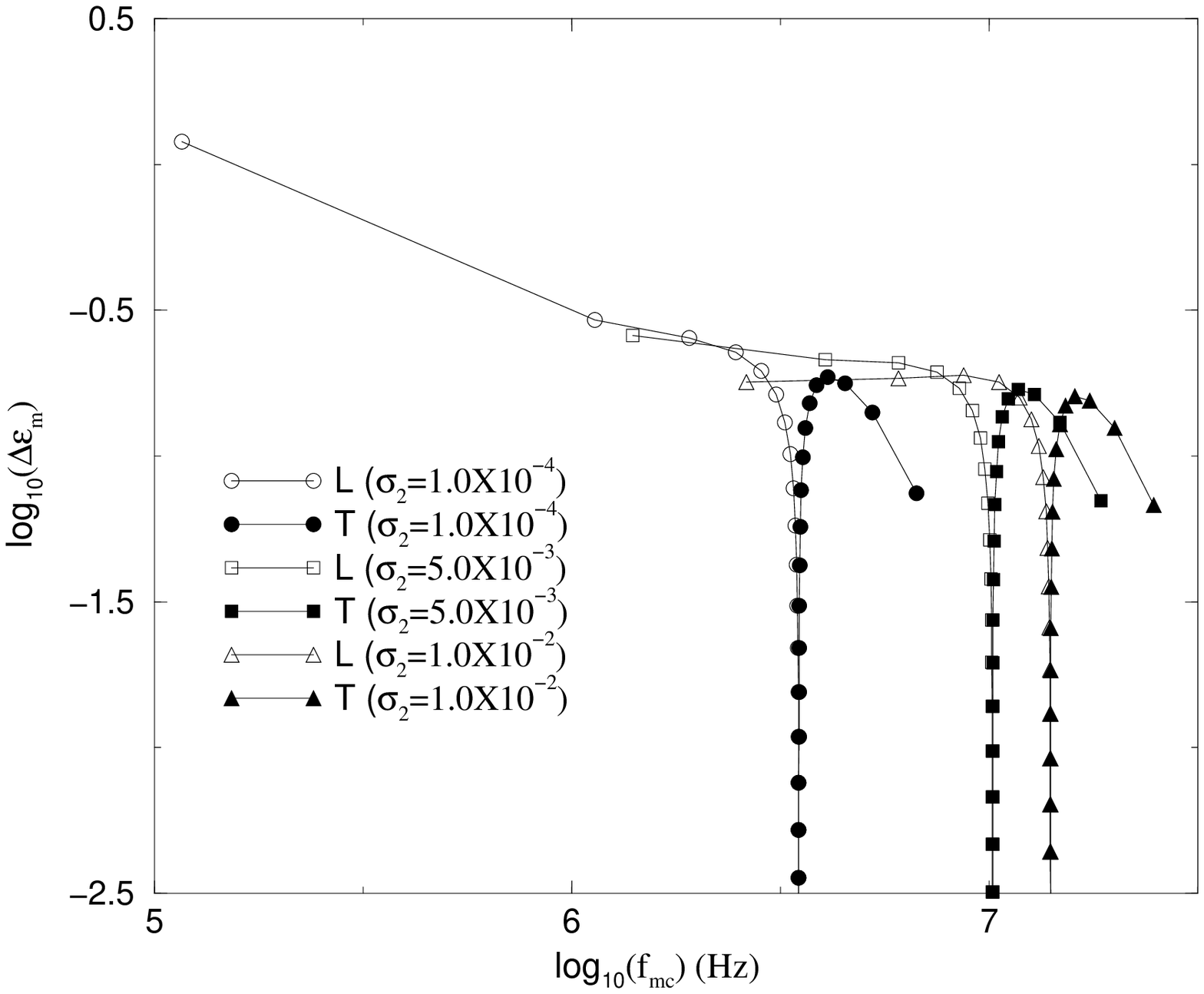,width=4.2in}
\caption{Dispersion strength versus the characteristic frequency 
for different medium conductivities. 
Parameters: $\zeta=0.12$V, $\eta=1.0\times 10^{-3}\,$Kg/(ms),  
$R/D=1.03$, $\ep_1=2.25\ep_0$, $u=0.033\,$C/m$^2$, 
$\Sigma=0.014\,$Sm$^2$/mol. The lines are drawn as a guide to the eye.}
\label{Fig.8.}
\end{figure}

Figure~\ref{Fig.2.} displays the effect of the  $\zeta$-potential.
It has been experimentally observed by Hughes and Green~\cite{JCIS} that
decreasing the $\zeta$-potential may red-shift the DEP crossover frequency. 
The system used by them  contained many latex beads
suspended in a solution, and hence the multipolar interaction
is expected to play a role. Our results
are in qualitative agreement with the above experimental findings. 
Furthermore, an increase in  the  $\zeta$-potential leads to 
higher $f_{\mathrm{CF}}$ in both the longitudinal and transverse field
cases. Similarly, increasing the real part of the dielectric constant leads to an 
increase in $f_{\mathrm{CF}}$, as displayed in Fig.~\ref{Fig.5.}. 
Increasing the viscosity of the medium (figure not shown here), however, 
has exactly the opposite effect for both the longitudinal and transverse field
cases. 

Figure~\ref{Fig.6.} shows the effect of  molar conductivity $\Sigma$ 
on crossover frequency.  
For small medium conductivities (here, $\sigma_2<10^{-2}\,$S/m), 
increasing $\Sigma$ leads to an increase
in the crossover frequency. However, there is a crossover 
after which  lower values of  $\Sigma$ yield higher
$f_{\mathrm{CF}}$. Similar behavior for the low surface conductivity
regime has been observed in experiments~\cite{JCIS}, but the authors
are not aware of any systematic study of the molar conductivity
on $f_{\mathrm{CF}}$. 
As Fig.~\ref{Fig.6.} shows, the effect is similar for both
longitudinal and transverse fields.

Figure~\ref{Fig.4.} shows the effect of varying the surface charge density on
the crossover frequency. Variations in the surface charge density lead to
more pronounced effects in the low frequency region, but close to the peak
the variations differences are very small. In addition, the location of 
the peak is only weakly dependent on surface charge density. These
results are in agreement with the experimental observations of 
Green and Morgan~\cite{green99a}.

In Fig.~\ref{Fig.7.}, we investigate the real part of the dipole
factor, and thus the DEP force.
The figure shows that
the effect due to multiple image plays an important role at low frequency region
when the particles separation is not large, whereas its
effect is smaller in the high frequency region. In the low
frequency region, the DEP force is be enhanced (reduced) due to the
presence of multiple images for longitudinal (transverse) field
case. As the particle separation grows,
the multiple image effect becomes negligible as
expected. We also studied the effect of particle size on the
real part of the dipole factor and the effect of multiple images increases
as the particle size decreases, and the effect is stronger in the longitudinal
field case.

Finally,  in Fig.~\ref{Fig.8.}, we plot the dispersion strengths ($\Delta\ep_m^{(L)}$ and
$\Delta\ep_m^{(T)}$) as a function of the characteristic frequencies ($f_{mc}^{(L)}$
and $f_{mc}^{(T)}$), for $m=1$ to $100$ with different medium
conductivities $\sigma_2$. Here, $m$ is a positive integer, 
and $F_m$ and $s_m$ are the microstructure parameters of the composite material,
see Eqs.~\ref{eq:fm}-\ref{eq:fmc}.
Hence, $\Delta\ep_m$ and $f_{mc}$ are the m-th
dispersion strength and characteristic frequency due to the presence of
multiple images as discussed in Sec.~\ref{sec:forma}.

The advantage of using the spectral representation theory is shown in
Fig.~\ref{Fig.8.}. Based on Fig.~\ref{Fig.7.}, it may appear that only one dispersion exist. 
Figure~\ref{Fig.8.} shows, however, that sub-dispersions with strength $\Delta\ep_m$ and
characteristic frequency $f_{mc}$ co-exist, and most of them lie close to the main dispersion.
Thus, the spectral representation theory helps us to gain 
more detailed information about the system and it provides a detailed comparison 
between the longitudinal and transverse field cases.

At a given $\sigma_2$, for the longitudinal (transverse) field
case, increasing $m$ leads to corresponding sub-dispersions in
the characteristic frequency due to the presence of multiple images.
The crossover frequencies $f_c$  are
$3.49\times 10^6\,$Hz, $1.0\times 10^7\,$Hz and $1.4\times 10^7\,$Hz (Fig.~\ref{Fig.8.}). 
From Fig.~\ref{Fig.8.} we find that at a lower medium
conductivity (say, $\sigma_2=1.0\times 10^{-4}\,$S/m), multiple images
have a stronger  effect on the DEP spectrum for the longitudinal field case
than for the transverse field. This is also apparent in  Fig.~\ref{Fig.7.} as well. 
Moreover, for longitudinal field case the multiple images play a role in the low frequency
range (i.e., smaller than $f_c$). For the transverse field the situation is the opposite.
At a larger $\sigma_2$ (say, $\sigma_2=5.0\times
10^{-3}\,$S/m or $\sigma_2=1.0\times 10^{-2}\,$S/m ), the
sub-dispersion strengths for the two cases have only a minor difference.
These observations may partly explain the results of Green and Morgan~\cite{green99a}
whose data suggests that there exists a dispersion below the frequencies
predicted by the current theory. The importance of these observation lies
in the fact that they help to clarify the interesting question of which
polarization mechanisms are present.

\section{Discussion and conclusion}
\label{sec:concl}

In this study, we have investigated the 
crossover spectrum of two approaching polarizable particles in the
presence of a nonuniform AC electric field. When the two particles
approach, the mutual polarization interaction between the particles
leads to changes in the dipole moments of each of the individual 
particles, and hence in the DEP crossover spectrum. This can
be interpreted as a correlation effect analogous to the ones
seen in charged systems~\cite{grosberg:02a}.

For charged particles, there is a  coexistence of
an electrophoretic and a dielectrophoretic force 
in the presence of a nonuniform AC electric
field. 
The DEP force always points toward the region
of high field gradient. It does not oscillate with the 
change of direction of the field. In contrast, the electrophoretic force points along the
direction of field, and hence is oscillatory under the same conditions. 
How to separate the DEP force from the electrophoretic force is a
question of interest in many experimental setups~\cite{Chou,bruckbauer02a}. 
In different frequency ranges, either the electrophoretic force or the DEP force
dominates, and the transition from one to the other occurs at a
frequency $f_{tr}$, which has been approximately determined~\cite{Mor1}. 
Here, we have chosen  a frequency region where electrophoretic effects 
are negligible and the DEP force dominates.
In addition, although we are at finite temperature, Brownian motion is 
not included in our analysis. In experiments Brownian motion is always
present and has posed difficulties in dielectrophoresis
of submicrometer particles. However, with current techniques it
is possible to access also this range~\cite{Mor1,marquet02a}.

One of the interesting questions is what happens, 
when the volume fraction of the suspension becomes large.
It turns out that it is possible to extend our approach 
by taking into account local field effects which may
modify the DEP crossover spectrum. Work is in progress
to address these questions. In addition to dielectrophoresis,
the extension of the present approach is also of 
interest from the point of view of electrorotation.

To summarize, using the  multiple image method, we have been able
to capture mutual polarization effects of two approaching 
particles in an electrolyte. Using spectral representation theory, we
derived an analytic expression for the DEP force, and using that the crossover
frequency was determined. From the theoretical analysis, we find that
the mutual polarization effects can change the crossover frequency
substantially. 

\acknowledgments
This work has been supported by the Research Grants Council of the Hong
Kong SAR Government under project number CUHK 4245/01P, and by the
Academy of Finland Grant No.~54113 (M.\,K.). J.P.H. is grateful to
Prof. K. Kaski for helpful discussions.

%%\bibliographystyle{apsrev}
%%\bibliography{dep}

%%%%%%%%%%%%%%%%%%%%%%%%%%%%%%%%%%%%%%%%%%%%%
%%%%%%%%%%%%%%%%%%%%%%%%%%%%%%%%%%%%%%%%%%%%%%

\end{document}